\documentclass[prb,twocolumn,showpacs,preprintnumbers,amsmath,amssymb,superscriptaddress, bibnotes]{revtex4}

\usepackage{color}
\usepackage{graphicx}
\usepackage{dcolumn}
\usepackage{bm}

\begin{document}

\title{Non-magnetic pair-breaking effect on La(Fe$_{1-x}$Zn$_{x}$)AsO$_{0.85}$ studied by NMR and NQR}

\author{S.~Kitagawa}
\email{shunsaku@scphys.kyoto-u.ac.jp}
\author{Y.~Nakai}
\author{T.~Iye}
\author{K.~Ishida}
\affiliation{Department of Physics, Graduate School of Science, Kyoto University, Kyoto 606-8502, Japan}
\affiliation{TRIP, JST, Sanban-cho bldg., 5, Sanban-cho, Chiyoda, Tokyo 102-0075, Japan}

\author{Y.~F.~Guo}
\author{Y.~G.~Shi}
\affiliation{International Center for Materials Nanoarchitectonics (MANA), National
Institute for Materials Science, Tsukuba, Ibaraki 305-0044, Japan}
\affiliation{TRIP, JST, Sanban-cho bldg., 5, Sanban-cho, Chiyoda, Tokyo 102-0075, Japan}

\author{K.~Yamaura}
\affiliation{TRIP, JST, Sanban-cho bldg., 5, Sanban-cho, Chiyoda, Tokyo 102-0075, Japan}
\affiliation{Superconducting Materials Center, National Institute for Materials
Science, 1-1 Namiki, Tsukuba, 305-0044 Ibaraki, Japan}

\author{E.~Takayama-Muromachi}
\affiliation{International Center for Materials Nanoarchitectonics (MANA), National
Institute for Materials Science, Tsukuba, Ibaraki 305-0044, Japan}
\affiliation{TRIP, JST, Sanban-cho bldg., 5, Sanban-cho, Chiyoda, Tokyo 102-0075, Japan}
\affiliation{Superconducting Materials Center, National Institute for Materials
Science, 1-1 Namiki, Tsukuba, 305-0044 Ibaraki, Japan}

\date{\today}

\begin{abstract}
$^{75}$As and $^{139}$La NMR and nuclear quadrupole resonance (NQR) studies on Zn-substituted LaFeAsO$_{0.85}$ have been performed to investigate the Zn-impurity effects microscopically. 
Although superconductivity in LaFeAsO$_{0.85}$ disappears by 3\% Zn substitution, we found that NMR/NQR spectra and NMR physical quantities in the normal state are hardly changed, indicating that the crystal structure and electronic states are not modified by Zn substitution. Our results suggest that the suppression of superconductivity by Zn substitution is not due to the change of the normal-state properties, but due to strong non-magnetic pair-breaking effect to superconductivity.  
\end{abstract}

\pacs{76.60.-k,	
74.25.-q, 
74.70.Xa 
}

\abovecaptionskip=-5pt
\belowcaptionskip=-10pt

\maketitle

 
The most important issues in iron-pnictide superconductors are to identify their superconducting (SC)-gap structure and pairing symmetry experimentally. Until now, it has become clear that iron-pnictide superconductors have a non-universal SC structure, i.e. various experiments suggest that $R$FeAs(O,F) ($R$:rare-earth elements) with the ``1111'' structure and (Ba,K)Fe$_2$As$_2$ with the ``122'' structure possess multi-finite SC gaps, but BaFe$_2$(As,P)$_2$ and Ba(Fe,Co)$_2$As$_2$ with heavily doping possess nodes in the SC gaps.\cite{J.Paglione_Naturephys_2010} At present, it remains controversial to identify the pairing symmetry and SC mechanism, because the SC mechanism is closely related with the gap function.       

On the theoretical side, soon after the discovery of the iron-pnictide superconductors, Mazin {\it et al.}, Kuroki {\it et al.} and Cvetkovic {\it et al.} independently proposed that spin fluctuations arising from nesting between the hole and electron Fermi surfaces (FSs) might give rise to the $s_{\pm}$-wave superconductivity with sign-reversing SC gaps.\cite{I.I.Mazin_PRL_2008,K.Kuroki_PRL_2008,V.Cvetkovic_EPL_2009} Various experimental results, e.g., observation of a resonance peak below $T_{\rm c}$ with neutron scattering measurements\cite{D.S.Inosov_Naturephys_2010} and the absence of a coherence peak in $1/T_1$ (nuclear spin-lattice relaxation rate) just below SC transition temperature $T_{\rm c}$,\cite{Y.Nakai_JPSJ_2008} appears consistent with the $s_{\pm}$-paring state.

In general, impurity responce of $T_{\rm c}$ gives a clue for SC properties.\cite{M.Sato_JPSJ_2010,Y.Li_NJP_2010,Y.F.Guo_PRB_2010,Y.Li_JPCS_2009,F.Hammerath_PRB_2010, Y.Nakajima_PRB_2010, M.Tropeano_PRB_2010,A.E.Karkin_PRB_2009}
Although the $s_{\pm}$-pairing is plausible from the theoretical and some experimental points of view, Sato {\it et al} threw doubt on the $s_{\pm}$-paring state on the basis of their experimental studies on La(Fe$_{1-y}$Co$_y$)As(O$_{1-x}$F$_x$) and Nd(Fe$_{1-y}$Ru$_y$)As(O$_{1-x}$F$_x$).\cite{M.Sato_JPSJ_2010} They claimed that $T_{\rm c}$ suppression rate by Co and Ru substitution is much smaller than the rate expected for superconductors with opposite signs of the order parameters. Following this stream, Onari and Kontani have theoretically proposed the s$_{++}$-wave symmetry mediated by orbital fluctuations.\cite{H.Kontani_PRL_2010}
In addition, Hammerath et al showed that $T_{\rm c}$ unexpectedly increase by the As-deficency in La1111 and suggested that pairing state might be modified by the As-deficency.\cite{F.Hammerath_PRB_2010}
This indicates that impurity effect for iron-pnictide superconductors is not so simple.

Recently, Nakamura {\it et al.} have shown from first-principle calculations that various transition-metal impurity effect needs careful theoretical treatments.\cite{K.Nakamura_arXiv_2010} They showed that impurity-$3d$ levels of substituted Co and Ni are close to the Fe-$3d$ level, so that the impurity effect is weak and the substitution dependence can be understood as a rigid band shift of the electronic structure. This is consistent with the experimental tendency.\cite{P.Canfield_PRB_2009} In addition, they pointed out that the Zn-$d$ level is considerably deep, resulting in that the Zn site can be regarded as a simple vacancy without effective carrier doping. Therefore, Zn atoms are considered as ideal impurity for studying impurity effects. However, Zn-impurity effects reported in La1111 superconductors are controversial: substituted Zn does not suppress superconductivity in under-doped and optimally-doped LaFeAs(O$_{1-x}$F$_x$),\cite{Y.Li_NJP_2010} but significantly suppress $T_{\rm c}$ in LaFeAsO$_{1-\delta}$\cite{Y.F.Guo_PRB_2010} and La(Fe$_{0.925}$Co$_{0.075}$)AsO.\cite{Y.Li_JPCS_2009} This discrepancy may originate from sample quality, and thus Zn substituted samples with careful characterization are required.

Here, we study the Zn-substitution effect in well-characterized LaFeAsO$_{1-\delta}$, in which superconductivity is completely suppressed by 3\% Zn substitution.\cite{Y.F.Guo_PRB_2010}
We have performed NMR and nuclear quadrupole resonance (NQR) measurements in order to identify the origin of the $T_{\rm c}$ suppression by Zn from the microscopic point of view.
From our NMR/NQR measurements, we found that Zn impurities do not change crystal structure nor normal-state electronic structure significantly, and thus the strong suppression of  $T_{\rm c}$ is considered to be due to non-magnetic pair-breaking effect, which cannot be interpreted as conventional $s$-wave superconductivity.

\begin{figure}[tb]
\vspace*{-10pt}
\begin{center}
\includegraphics[width=9cm,clip]{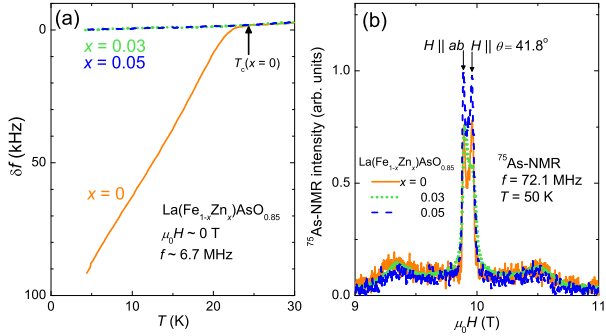}
\end{center}
\caption{(Color online) (a) Meissner signals for La(Fe$_{1-x}$Zn$_{x}$)AsO$_{0.85}$ ($x = 0, 0.03, {\rm and}~0.05$) measured by an identical NMR coil. The arrow indicates $T_{\rm c}$ at $x = 0$.
(b) $^{75}$As NMR spectra for $x = 0, 0.03, {\rm and}~0.05$ at 50~K obtained by sweeping magnetic field at a fixed frequency of 72.1~MHz. Compared with Zn-free LaFeAsO$_{0.85}$, neither significant broadening nor additional peaks were observed in the Zn-substituted samples.}
\label{Fig.1}
\end{figure}

Polycrystalline samples of Zn-substituted La(Fe$_{1-x}$Zn$_{x}$)AsO$_{0.85}$ ($x$ = 0, 0.03, and 0.05) synthesized by solid-state reaction under high pressure\cite{Y.F.Guo_PRB_2010} are ground into powder for our NMR/NQR measurements. 
$T_{\rm c}$ was determined from the onset temperature of Meissner signal measured by an identical NMR coil.
 $T_{\rm c}$ of Zn-free $x = 0$ is 24~K, and the $x$ = 0.03 and 0.05 samples do not show any Meissner signals as shown in Fig.~\ref{Fig.1} (a). This is consistent with the previous report.\cite{Y.F.Guo_PRB_2010} 
Conventional spin-echo technique was utilized for following NMR/NQR measurements. 

Figure \ref{Fig.1} (b) shows $^{75}$As NMR spectra for $x$ = 0, 0.03, and 0.05 in La(Fe$_{1-x}$Zn$_{x}$)AsO$_{0.85}$ at 50 K obtained by sweeping magnetic field at a fixed frequency of 72.1~MHz. 
Compared with Zn-free $x = 0$, neither significant broadening nor additional peaks were observed in the Zn-substituted samples.
\begin{figure}[tb]
\vspace*{-10pt}
\begin{center}
\includegraphics[width=9cm,clip]{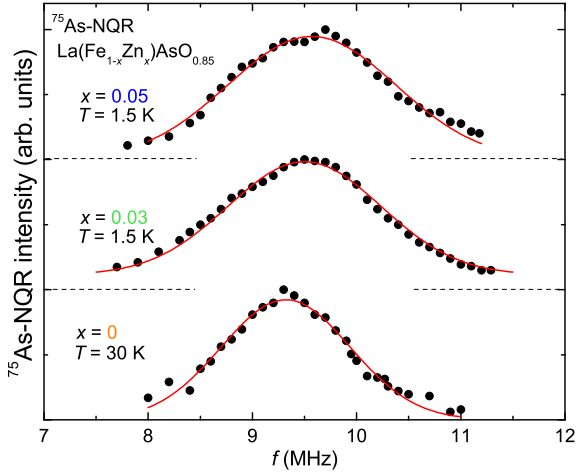}
\end{center}
\caption{(Color online) $^{75}$As NQR spectra for $x = 0$ at 30~K, 0.03 and 0.05 at 1.5~K. Solid lines represent Gaussian fits.}
\label{Fig.2}
\end{figure}
Figure \ref{Fig.2} shows the $^{75}$As NQR spectra for each sample, which were obtained by frequency-swept method. The peak frequencies ($\nu_{\rm Q}$) of the $^{75}$As NQR spectra slightly increase, and the line width of each sample becomes broadened by Zn substitution, indicative of randomness of the electric field gradient introduced by Zn substitution. 

\begin{figure}[tb]
\vspace*{-10pt}
\begin{center}
\includegraphics[width=9cm,clip]{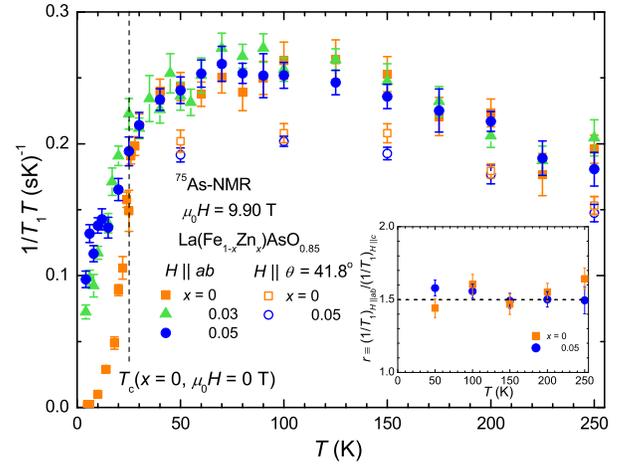}
\end{center}
\caption{(Color online) (Main panel) Temperature dependence of 1/$T_1T$ for $x = 0, 0.03, {\rm and}~0.05$. 1/$T_1T$ in the normal state hardly changes by Zn substitution.
(Inset) Temperature dependence of the anisotropy of 1/$T_1T$, $r \equiv$ (1/$T_1$)$_{H\parallel ab}$/(1/$T_1$)$_{H\parallel c}$, for $x = 0$ and 0.05. (1/$T_1$)$_{H\parallel c}$ is estimated from (1/$T_1$)$_{H\parallel ab}$ and (1/$T_1$)$_{H\parallel 41.8^{\rm \circ}}$. (1/$T_1$)$_{H\parallel ab}$/(1/$T_1$)$_{H\parallel c}$ $\simeq$ 1.5 does not change by Zn substitution. }
\label{Fig.3}
\end{figure}
Figure \ref{Fig.3} shows temperature dependence of 1/$T_1T$ for $x$ = 0, 0.03, and 0.05. 
$1/T_{1}$ was measured in $\mu_0 H \simeq 9.9$~T for $H \parallel ab$.
On cooling, 1/$T_1T$ at $x = 0$ slightly increases down to 100~K, and decreases from $\sim 50$~K above $T_{\rm c}$. From comparing with previous NMR/NQR results reported by Mukuda {\it et al.},\cite{H.Mukuda_PhysicaC_2009} electron doping of our $x = 0$ sample is lower, since $1/T_1T$ at 100 K is larger and the NQR frequency ($\nu_Q \sim 9.3$ MHz) is lower than those in their underdoped sample. 
In the $x = 0.03$ and $0.05$ samples, 1/$T_1T$ continues to decreases through 50~K. Since SC Meissner signal was not observed, the decrease of 1/$T_1T$ is not due to superconductivity, but can be ascribed to the characteristic band dispersion around the Fermi energy.\cite{H.Ikeda_JPSJ_2008,H.Ikeda_PRB_2010} 
The pseudogap like behavior observed in 1/$T_1T$ can be interpreted by the presence of high density of states just below the Fermi energy, and continues down to low temperatures in the Zn-substituted samples.
However, 1/$T_1T$ above 30~K hardly changes by Zn substitution, suggesting that Zn substitution modifies neither carrier doping level nor FS properties.

Although we tried to align the samples as we performed previously in LaFeAs(O$_{1-x}$F$_x$) to investigate the anisotropy of $1/T_1$,\cite{S.Kitagawa_PRB_2010} we could not align them. Alternatively, we measured $1/T_1$ at the peak corresponding to $H \parallel \theta = 41.8^{\circ}$ [shown in Fig.~\ref{Fig.1} (b)] to derive $1/T_1$ along the $c$-axis. Here, $\theta$ is an angle between a magnetic field and the principal axis of the electric field gradient ($c$-axis). In general, the angle dependence of $1/T_1$ in axial symmetric crystals can be described as 
\begin{align}
1/T_1 (\theta) = (1/T_1)_{H\parallel c} \cos^2\theta + (1/T_1)_{H\parallel ab} \sin^2\theta. 
\end{align}
Therefore, (1/$T_1$)$_{H \parallel c}$ and the anisotropy of 1/$T_1$ [$r \equiv  (1/T_1)_{H \parallel ab} / (1/T_1)_{H \parallel c}$] can be estimated from (1/$T_1$)$_{H \parallel ab}$ and 1/$T_1$ measured at $\theta = 41.8^{\circ}$ by using the above relation.
The inset of Fig.~\ref{Fig.3} shows the temperature dependence of $r$. The value of  $r$ is approximately 1.5 at $x =$ 0 and 0.05 above 50 K, suggesting that the local stripe correlations are unchanged by Zn substitution, since $r \simeq 1.5 $ can be understood by the presence of the local stripe correlations related with the nesting between the hole and electron FSs.\cite{K.Kitagawa_JPSJ_2009,S.Kitagawa_PRB_2010}


Now, we discuss the Zn-substitution effect in LaFeAsO$_{0.85}$. Several effects which suppress $T_{\rm c}$ can be pointed out; variations of (i) crystal structure and (ii) electronic structure, and (iii) induction of staggered magnetism by nonmagnetic impurities.
It was reported that the Fe-As-Fe bond angle and/or pnictogen height are important parameters for determination of $T_{\rm c}$.\cite{Y.Mizuguchi_SST_2010, C.-H.Lee_JPSJ_2008} NQR studies on LaFeAs(O$_{1-x}$F$_{x}$) and LaFeAsO$_{1-\delta}$ have shown that $\nu_{\rm Q}$ is related to the hybridization between the Fe and As orbitals, and thus related to the As-Fe-As bond angle.\cite{H.Mukuda_PhysicaC_2009, S.Kitagawa_PhysicaC_2010, G.Lang_PRL_2010} With increasing F content in LaFeAs(O$_{1-x}$F$_{x}$), $\nu_{\rm Q}$ in LaFeAs(O$_{1-x}$F$_{x}$) increases by 1.5~MHz and As-Fe-As bond angle increases by 0.8$^\circ$ from undoped to overdoped samples.\cite{Q.Huang_PRB_2008,S.Kitagawa_PhysicaC_2010} 
Following this relation, observed $\nu_{\rm Q}$ change of 0.28~MHz by 5\% Zn substitution corresponds to 0.15$^\circ$ increase of As-Fe-As bond angle, which is consistent with the XRD result.\cite{Y.F.Guo_PRB_2010}
The tiny change in $\nu_{\rm Q}$ with Zn substitution indicates that the hybridization between the Fe and As bonds is almost unchanged, and thus the $T_{\rm c}$ suppression by Zn substitution cannot be attributed to variations of the Fe-As-Fe bond angle and/or pnictogen height.

Next, we consider the variations of electronic state from the viewpoint of low-energy magnetic fluctuations. 
In LaFeAs(O$_{1-x}$F$_x$) and LaFeAsO$_{1-\delta}$, low-energy magnetic fluctuations probed with $1/T_1$ measurements are dramatically suppressed by F (electron) doping due to variations of the nesting condition.\cite{Y.Nakai_NJP_2009,H.Mukuda_JPSJ_2009,T.Nakano_PRB_2010} Therefore, the variations of the electronic state with Zn substitution should be detected with $1/T_1$ measurements.
As seen in Fig.~\ref{Fig.3}, $1/T_1T$ in the normal state remains unchanged, indicating that Zn substitution does not modify the FS properties. This result is in good agreement with the Hall-coefficient and specific-heat results, in which normal-state data are identical between Zn-free and substituted samples.\cite{Y.F.Guo_PRB_2010}
Moreover, the stripe antiferromagnetic (AF) spin correlations, which originate from the nesting between electron and hole FSs are essentially unchanged, since the anisotropy of $1/T_1$ is the same in the $x$ = 0 and 0.05 samples.\cite{S.Kitagawa_PRB_2010}

\begin{figure}[tb]
\vspace*{-10pt}
\begin{center}
\includegraphics[width=9cm,clip]{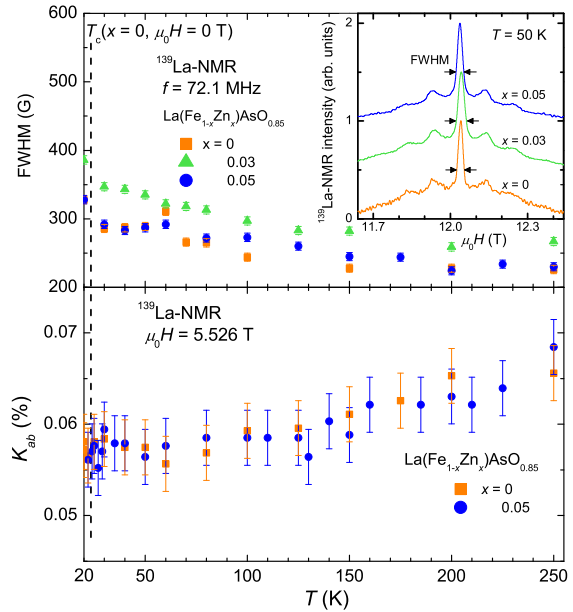}
\end{center}
\caption{(Color online) (Upper panel) Temperature dependence of the full width at half maximum (FWHM) of $^{139}$La NMR spectra for $x = 0, 0.03, {\rm and}~0.05$. All samples show similar behavior.
(Inset) Field swept $^{139}$La NMR spectra at 50 K at 72.1 MHz.
(Lower pannel) Temperature dependence of $^{139}$La NMR Knight shift for $H\parallel ab$.}
\label{Fig.4}
\end{figure}
Magnetism potentially induced by non-magnetic impurities is also investigated by $^{139}$La NMR measurements. Since the hyperfine coupling constant at the La site is smaller than at the As site, the wipe out effect due to induced magnetism is expected to be weak at the La site even if it exists. Figure \ref{Fig.4} shows temperature dependence of the full width at half maximum (FWHM) of the central peaks shown in the inset of Fig.~\ref{Fig.4} and the Knight shift derived from $^{139}$La NMR spectra. 
The FWHM of all samples becomes broader with decreasing temperature,
but the temperature dependence in the normal state is nearly the same, indicating that this line-width broadening does not originate from Zn substitution but from extrinsic magnetic impurities present even in the Zn-free sample.
Furthermore, $^{139}$La NMR Knight shift is identical between the $x = 0$ and $x =$ 0.05 samples, and does not show any Curie-Weiss behavior, as shown in Fig.~\ref{Fig.4}.
These results indicate that the substituted Zn impurity does not induce any local moments, in quite contrast with Zn-substituted cuprates, particularly underdoped cuprates. It is well known that non-magnetic impurities in underdoped cuprates induce the staggered AF moments around substituted impurities, which are explicitly observed by $^{89}$Y NMR in YBa$_2$(Cu$_{0.99}$Zn$_{0.01}$)O$_{6.64}$ and $^{27}$Al NMR in La$_{1.85}$Sr$_{0.15}$(Cu$_{0.97}$Al$_{0.03}$)O$_4$.\cite{A.Mahhajan_PRL_1994,K.Ishida_PRL_1996} 
The difference between the Zn-substitution effects of LaFeAsO$_{1-\delta}$ and of under-doped cuprates presumably originates from the different nature of magnetism in parent compounds and mobility of carriers.
The underdoped cuprates possess strong low-energy AF correlations, and substituted Zn impurities destroy the AF correlations and induce local moments around Zn impurities. In contrast, the parent compounds of iron pnictides are interpreted to be of itinerant nature and the carrier mobility is higher than that of cuprates, thus substituted Zn does not induce local moments as in conventional metallic compounds.  

The important question to be clarified is why only 3\% Zn substitution suppresses superconductivity completely in LaFeAsO$_{1-\delta}$. The electron localization observed in Zn-substituted underdoped cuprates was suggested as the origin of the $T_{\rm c}$ suppression,\cite{M.Sato_JPSJ_2010} but this possibility can be excluded since the substituted Zn neither modifies electronic structures, nor induces staggered antiferromagnetism. The strong $T_{\rm c}$ suppression by non-magnetic Zn impurities cannot be interpreted by the conventional $s$-wave superconductivity, but strongly suggests unconventional nature of superconductivity in LaFeAsO$_{1-\delta}$. 

In the iron-pnictide superconductors, $s_{\pm}$ wave superconductivity is regarded as a promising SC state, and its local impurity effect was studied based on the five-orbital model.\cite{S.Onari_PRL_2009} In this model, superconductivity is expected to vanish when $g > g^{s\pm}_{\rm c} = 0.23$ where $g = z\gamma /2\pi T_{c0}$ is a pair-breaking factor. The critical concentration of superconductivity was estimated to be a few \%, depending on substituted impurity potential, which is in good agreement with the present Zn impurity effect. However, if one estimates $g$ by using experimental values of residual resistivity, a $g$ value becomes extremely large, resulting in the conclusion that the impurity effects are negligibly small or absent in La1111 superconductors. We point out that the residual resistivity in polycrystalline samples might not reflect intrinsic impurity effect properly, but mainly reflect grain boundary effect, since the systematic variation of residual-resistivity value is difficult to be observed experimentally in polycrystalline La1111 with impurities. 

Finally, we comment on the previous reports on the Zn-substitution effect in LaFeAs(O$_{1-x}$F$_x$).\cite{Y.Li_NJP_2010} In the previous reports, the normal-state resistivity in Zn-substituted samples is smaller than that in the Zn-free samples, which is quite unusual and different from the results of our samples.\cite{Y.F.Guo_PRB_2010} The decrease of the normal-state resistivity by Zn substitution is hard to understand because substituted Zn does not change carrier content as clarified in this paper. We point out that the carrier content might be changed in the Zn-substitution process, which can be checked by $^{75}$As NMR/NQR measurements as shown in this paper. We claim that microscopic NMR and NQR studies play a crucial role to discuss the impurity effect in the iron-pnictide superconductors.    
          
In conclusion, from the microscopic NMR and NQR measurements, we show that the Zn substitution neither changes carrier content nor modifies the electronic state in LaFeAsO$_{1-\delta}$. In addition, temperature dependence of $^{139}$La NMR spectra indicates that the substituted Zn does not induce any static moments around the impurity in contrast to underdoped cuprates. Superconductivity is suppressed by only a few percent of non-magnetic Zn substitution, which cannot be expected in conventional superconductors, but strongly suggests unconventional nature of superconductivity in LaFeAsO$_{1-\delta}$. The Zn-substitution effect revealed by the present study gives a strong constraint to theoretical models for the iron-pnictide superconductors.   

We are grateful to S. Yonezawa, H. Ikeda, and Y. Maeno for valuable discussions. 
This work was supported by a Grant-in-Aid for the Global COE Program ``The Next Generation of Physics, Spun from Universality and Emergence'' from MEXT, and for Scientific Research from JSPS.
This was also supported in part by ``Funding Program for World-Leading
Innovative R\&D on Science and Technology (FIRST Program)''.

\end{document}